\documentclass[
    ,final            
  ]
  {aipproc}

\layoutstyle{6x9}

\def\apj{{\it Astrophys. J.} }
\def\apjl{{\it Astrophys. J. Lett.} }
\def\apjs{{\it Astrophys. J. Suppl.} }
\renewcommand{\citep}[1]{\cite{#1}}
\renewcommand{\citealp}[1]{\cite{#1}}
\newcommand{\beq}{\begin{equation}}
\newcommand{\eeq}{\end{equation}}

\begin{document}

\title{A key to the spectral variability of prompt GRBs}


\author{Mikhail V. Medvedev}{
  address={Department of Physics and Astronomy, University of Kansas, 
Lawrence, KS 66045}
}

\begin{abstract}
We demonstrate that the rapid spectral variability of prompt GRBs is 
an inherent property of radiation emitted from shock-generated, 
highly anisotropic small-scale magnetic fields. We interpret the 
hard-to-soft evolution and the correlation of the soft index $\alpha$ 
with the photon flux observed in GRBs as a combined effect of temporal 
variation of the shock viewing angle and relativistic aberration of an 
individual thin, instantaneously illuminated shell. The model predicts 
that about a quarter of time-resolved spectra should have hard spectra, 
violating the synchrotron $\alpha=-2/3$ limit. The model also naturally 
explains why the peak of the distribution of $\alpha$ is at $\alpha\sim-1$. 
The presence of a low-energy break in the jitter spectrum at oblique angles 
also explains the appearance of a soft X-ray component in some GRBs and 
their paucity. We emphasize that our theory is based solely on the first 
principles and contains no ad hoc (phenomenological) assumptions.
\end{abstract}

\maketitle


\section{Introduction}

\begin{figure}
  \includegraphics[height=.2\textheight]{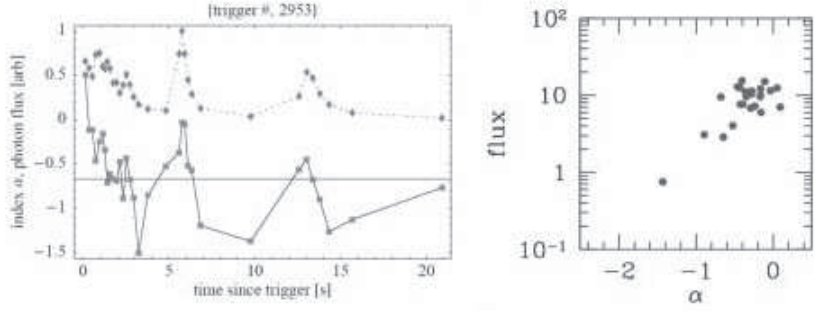}
  \caption{(a) --- A tracking GRB: normalized flux (diamonds) and the 
soft spectral index $\alpha$ (squares) evolve similarly with time. 
(b) --- Scatter plot of flux vs. $\alpha$ for GRB940429.    }
\label{f:1}
\end{figure}

Rapid spectral variability is a remarkable, yet unexplained feature of 
the prompt GRB emission. The variation of the hardness of the spectrum 
and the hard-to-soft evolution are the most acknowledged features 
\citep{Bhat+94,Crider+97,RP02}. A quite remarkable 
``tracking'' behavior, when the low-energy spectral index $\alpha$ 
follows (or correlates with) the photon flux \citep{Crider+97} is 
particularly intriguing. An example of such a trend is shown in 
Fig. \ref{f:1} (we used data from \citealp{Preece+00}).

\section{Theory of jitter radiation}

The angle-averaged spectral power emitted by a relativistic particle 
moving through small-scale Weibel-generated magnetic fields
is given here without derivation (see \cite{M00,M06} for details):

\beq
\frac{dW}{d\omega}=\frac{e^2\omega}{2\pi c^3}\int_{\omega/2\gamma^2}^\infty
\frac{\left|{\bf w}_{\omega'}\right|^2}{\omega'^2}
\left(1-\frac{\omega}{\omega'\gamma^2}+\frac{\omega^2}{2\omega'^2\gamma^4}
\right)\,d\omega' .
\label{dW/dw}
\eeq
Here $\gamma$ is the Lorentz factor of a radiating particle and 
${\bf w}_{\omega'}=\int{\bf w}e^{i\omega't}\,dt$ is the Fourier  
component of the transverse particle's acceleration due to the 
Lorentz force. This temporal Fourier transform is taken along the 
particle trajectory, ${\bf r}={\bf r}_0+{\bf v}t$.


\begin{figure}
  \includegraphics[width=.5\textheight]{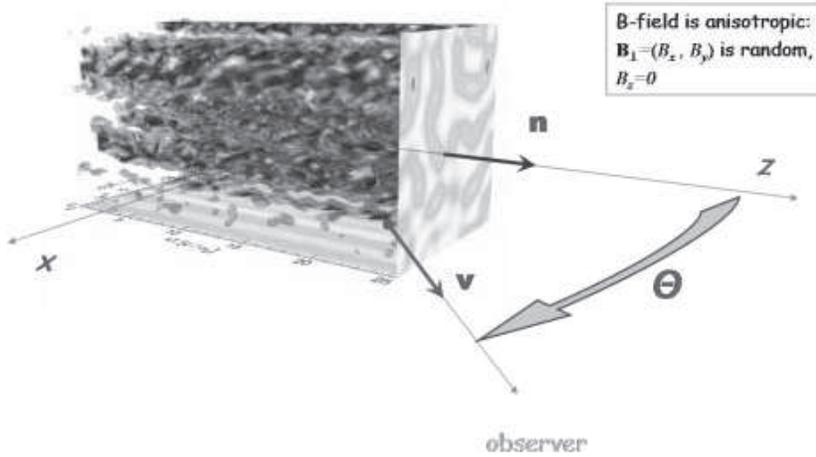}
  \caption{Magnetic filaments in a shock \cite{MSK06}. The radiation
spectrum varies with the viewing angle $\Theta$.}
\label{f:2}
\end{figure}

Because the Weibel-generated magnetic fields at a shock are highly 
anisotropic, as shown in Figure \ref{f:2}, the spectra of radiation emitted
by an electron in such fields depend on the viewing 
angle $\Theta$ between the normal to the shock and the particle velocity
in the shock co-moving frame, which is approximately the direction 
toward an observer, for an ultra-relativistic particle. 
The acceleration spectrum is 
\beq
\langle|{\bf w}_{\omega'}|^2\rangle
={C}/{2\pi}\,(1+\cos^2\Theta)\int\!\! f_z(k_\|) f_{xy}(k_\perp)
\delta(\omega'+{\bf k\cdot v})\,dk_\|d^2 k_\perp,
\label{w-main}
\eeq
where $C$ is the mormalization of the magnetic feld spectrum 
over spatial scales to the total field energy density $B^2/8\pi$.
Here $f_{xy}$ and $f_{z}$ determine spectra of the magnetic fields
in the shock plane and in the direction of the shock motion. These 
spectra are independent of each other to a large degree, as indicated 
by numerical simulation (e.g., \cite{Fred+04}). We assume  $f_{xy}$ 
and $f_{z}$ are broken double-power-laws,
\beq
f_z(k_\|)={k_\|^{2\alpha_1}}/{(\kappa_\|^2+k_\|^2)^{\beta_1}}, \quad 
f_{xy}(k_\bot)
={k_\perp^{2\alpha_2}}/{(\kappa_\perp^2+k_\perp^2)^{\beta_2}},
\label{f}
\eeq 
with the position of the break
(peak) determined by the Weibel scale: the plasma skin depth,
$\kappa_\bot\sim\kappa_\|\sim\omega_p/c$.

\section{Interpretation of prompt GRB spectra}

\begin{figure}
  \includegraphics[width=.3\textheight]{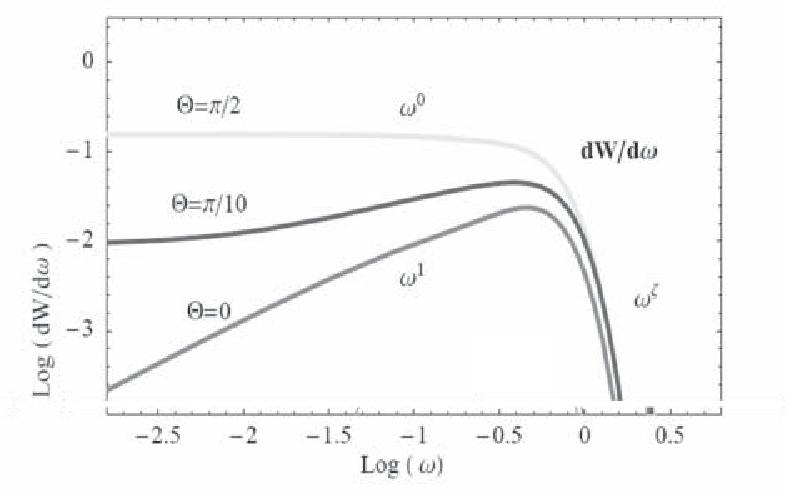}
  \includegraphics[width=.3\textheight]{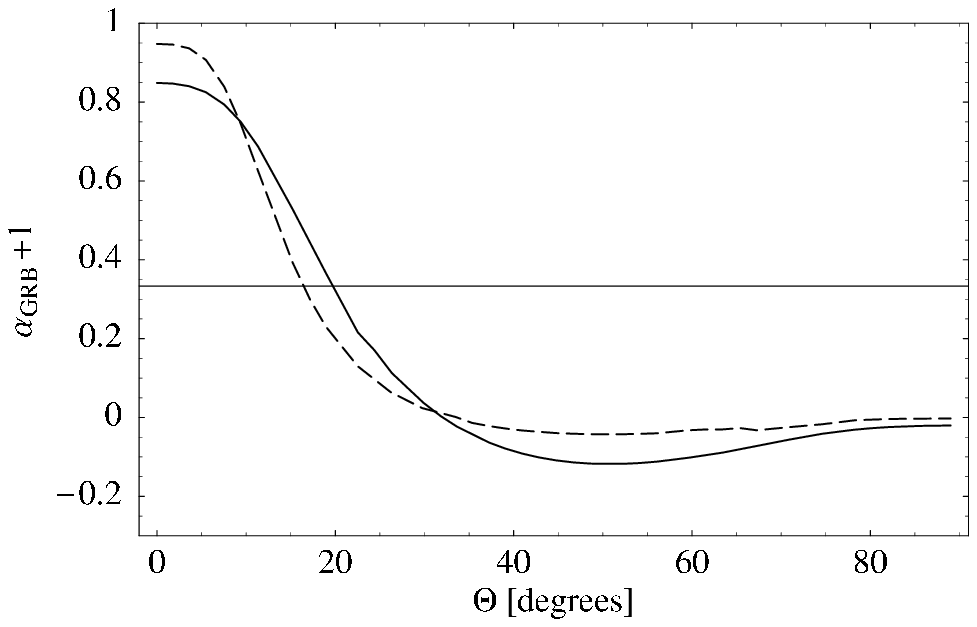}
  \caption{(a) --- Typical spectra for three viewing angles,
$\Theta=0,\ \pi/10,\ \pi/2$. (b) --- Soft sceltral index as a function
of $\Theta$ at energies 10 (solid) and 30 (dashed) times below $E_p$.}
\label{f:3}
\end{figure}

The typical jitter
radiation spectra for three different angles are shown in Figure \ref{f:3}a. 
 When a shock velocity is 
along the line of sight, the low-energy spectrum is hard 
$F_\nu\propto\nu^1$, harder than the ``synchrotron line of death'' 
($F_\nu\propto\nu^{1/3}$). As the viewing angle increases, the 
spectrum softens, and when the shock velocity is orthogonal to the 
line of sight, it becomes $F_\nu\propto\nu^0$. Another interesting 
feature is that at oblique angles, the spectrum does not soften 
simultaneously at all frequencies. Instead, there appears a smooth 
spectral break, which position depends on $\Theta$. The spectrum 
approaches $\sim\nu^0$ below the break and is harder above it. 
This softening of the spectrum at low $\nu$'s could be interpreted 
as the appearance of an additional soft X-ray component, similar to 
that found in some of GRBs \citep{Preece+95}. 

Figure \ref{f:3}b shown the low-energy slope at a frequency 10 and 30
times lower than the spectral peak.
These frequencies 
correspond to the edge of the {\it BATSE} window for bursts with 
the peak energy of about 200~keV and 600~keV, respectively. Hence, 
the spectral slope, $\alpha_{\rm GRB}$, will be close to those 
obtained from the data fits. Since $\Theta(t)$ increases with time 
during an individual emission episode, the curves roughly represent the 
temporal evolution of $\alpha_{\rm GRB}$. Assuming that 
time-resolved spectra are homogeneously distributed over $\Theta$, 
one can estimate the relative fraction of the synchrotron-violating 
GRBs (i.e., those with $\alpha_{\rm GRB}+1>1/3$) as about 25\%, 
which is very close to the 30\% obtained from the data \citep{Preece+00}. 
Most of the GRBs, $\sim$75\%, should, by the same token, be distributed 
around $\alpha_{\rm GRB}\sim -1$. Note also that time-integrated GRB 
spectra should have $\alpha_{\rm GRB}$ around minus one, as well.

\begin{figure}
  \includegraphics[width=.3\textheight]{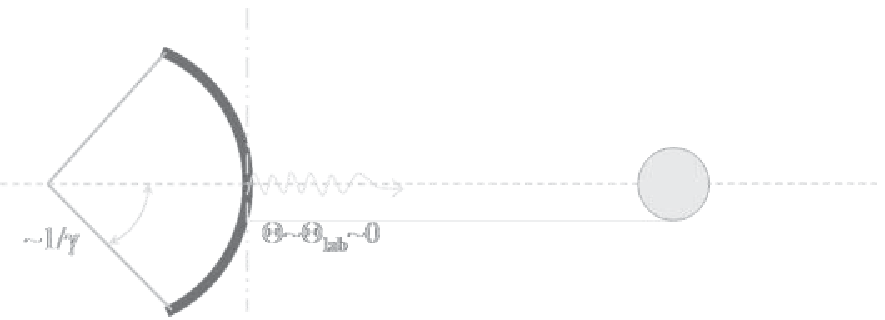}~~~~~
  \includegraphics[width=.3\textheight]{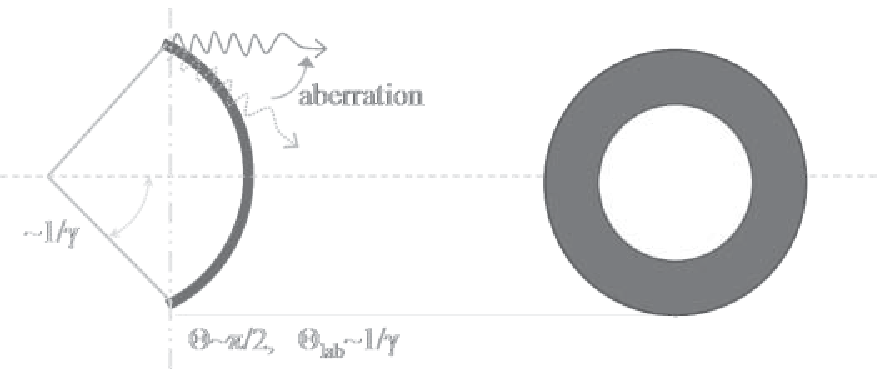}
  \caption{Cartoon explaining correlation of the spectral index and
the flux as a combined effect of the anisotropy of jitter radiation and 
relativistic aberration.}
\label{f:4}
\end{figure}

In the standard internal shock model, each emission episode is associated 
with illumination of a thin shell, --- an internal shock and the hot and 
magnetized post-shock material. We assume that the shell is spherical 
(at least within a cone of opening angle of $\sim1/\Gamma$ around the 
line of sight) and this shell is simultaneously illuminated for a short 
period of time. The observed photon pulse is broadened because the photons 
emitted from the patches of the shell located at larger angles, 
$\vartheta$, from the line of sight arrive at progressively later 
times. The bolometric flux depends on $\vartheta$, and hence on 
time as \cite{RP02}:
\beq
F_{\rm bol}=F_0{\cal D}^2(\Theta)/\Gamma^2
\eeq
with the Lorentz boost
\beq
{\cal D}(\Theta)=\left[\Gamma(1-\beta\cos\vartheta)\right]^{-1}
=\Gamma(1+\beta\cos\Theta)
=\left[\Gamma(1-\beta-\beta c \Delta t/R_0)\right]^{-1}.
\eeq
Because of relativistic 
aberration, the comoving viewing angle, $\Theta$, is greater 
than $\vartheta$ and approaches $\Theta\sim\pi/2$ (the shell is 
seen edge-on) when $\vartheta\sim1/\Gamma$. Thus, there must be a 
tight correlation between the observed spectrum and the observed 
photon flux, because they are, in essence, different manifestations 
of the same relativistic kinematics effect. An observer first detects
photons emitted close to the line of sight, for which 
$\vartheta\sim\Theta\sim0$, see Figure \ref{f:4}a. The flux is
the largest and $\alpha$ is the hardest (around $0$) at this time. 
As time goes on, an observer sees photons emitted farer from the line
of sight, as in Figure \ref{f:4}b. The flux decreases, 
whereas $\alpha$ becomes softer and approaches $-1$.
Of course, we neglected cooling effects here, which 
can result in even softer spectra with $\alpha\sim-3/2$.

\begin{theacknowledgments}
This work has been supported by NASA grant NNG-04GM41G, 
DoE grant DE-FG02-04ER54790, and the KU GRF fund. 
\end{theacknowledgments}

\bibliographystyle{aipproc}   


\end{document}